# THE PROPOSED NLC LINAC LLRF SYSTEM*

P. Corredoura†, C. Adolphsen
Stanford Linear Accelerator Center, Stanford, Ca 94309, USA

*Abstract*

The proposed Next Linear Collider (NLC) contains linac systems operating at L,S,C, and X band. This paper describes a wideband modular low-level RF (LLRF) system applicable for all NLC pulsed RF systems. High speed digital IF techniques are used for both arbitrary klystron drive modulation and accurate RF vector detection. Measurement of relative beam to RF phase, beam loading compensation, and structure dipole modes for automated alignment are all handled by the system. Modern signal processing techniques are also described.

## 1. INTRODUCTION

The NLC is essentially a pair of opposing X-band linacs designed to collide 500 GeV electrons and positrons. The main linac RF system is a major driving cost for the project consisting of 1600 X-band klystron tubes and related hardware. Pulse stacking RF techniques are used to efficiently develop the 600MW 300ns RF pulses required at each girder. A group of 8 klystrons (an 8 pack) deliver RF power to 8 girders, each girder supports 3 accelerator structures. Pulse stacking and beam loading compensation requires fast modulation of the klystron drive signal. A programmable digital IF solution will be presented.

The absolute phase of the accelerating RF with respect to the beam is a critical parameter which must be measured and controlled to the 1 degree X-band level. Even with a state-of-the-art stabilized fiber optic RF reference/distribution system, there will still be phase drifts present in the system which will require measuring the relative phase between the beam and accelerating RF at regular intervals. The techniques for making this and other X-band measurements are described in this paper can be applied to any linac RF system.

Transverse alignment must be achieved to extremely tight tolerances to prevent excitation of transverse modes. Each accelerator is a damped detuned structure which is designed to load down the undesirable high-order modes (HOMs) and allow their external detection to facilitate alignment of each girder [1]. The shape of each individual accelerating cell in a structure is altered slightly along the length of the structure making the HOM frequencies a function of the longitudinal position on the structure. The frequency of the lowest order transverse mode ranges from 14 to 16 GHz corresponding to upstream and downstream respectively [2]. A receiver will be outlined which measures the beam induced HOMs allowing automated alignment of each girder via remote mechanical movers.

*Work supported by Department of Energy, contract DE-AC03-76SF00515

† plc@slac.stanford.edu

## 2. KLYSTRON DRIVE GENERATION

For the main linac system the output of 8 klystrons will be combined to deliver X-band RF to 8 accelerator girders through a high power distribution and delay system (DLDS) (figure 1). Each ~3us klystron pulse will be time multiplexed to steer RF power successively to 8 accelerator girders by quickly modulating the relative phases of the klystron drive during a pulse. Wideband klystrons (>100MHz) will allow rapid phase switching to improve overall efficiency. Fixed delays in the DLDS deliver RF power a fill time in advance of the beam.

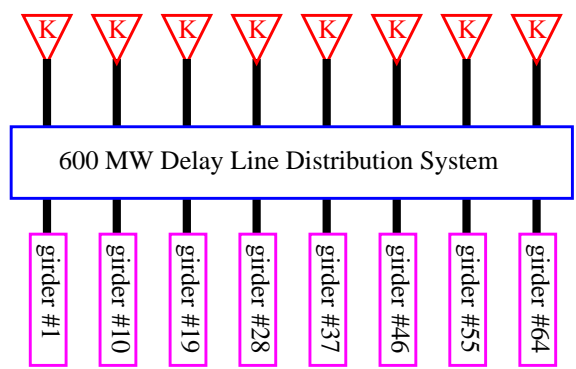

Fig. 1. Diagram showing 8 klystron driving 8 linac girders.

To compensate for beam loading, klystrons will be precisely phase ramped during the first ~100ns of each sub-pulse to direct some RF power to a port which is out of time with the beam. This allows the klystrons to operate in saturation while minimizing the energy spread across the 95 bunch train. To compensate for unknown transmission phase shifts the system must produce any drive phase and allow a smooth transition to any other phase. Specifications for the main linac LLRF drive system are listed below in table 1.

| Parameter | Value |
|---|---|
| carrier frequency | 11.424 GHz |
| pulse width | 3.1 us |
| bandwidth | >100 MHz |
| phase range | arbitrary, continuous |
| phase resolution | <1 degree |
| dynamic range | > 20 dB |

Table 1: Main linac LLRF drive requirements.

An in-phase/quadrature (IQ) drive generation system was produced for the next linear collider test accelerator (NLCTA) [3]. Two high speed (250 MS/S) DACs were used to drive an X-band baseband IQ modulator (figure 2). The IQ technique works but is sensitive to mixer offsets, quadrature errors, baseband noise and requires two DACs.

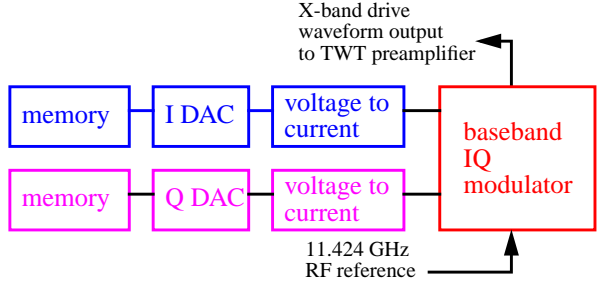

Fig. 2. Baseband IQ technique used to produce the X-band klystron drive waveform in the NLCTA.

To reduce system cost and improve accuracy a digital IF approach is being pursued (figure 3). A modulated IF tone burst is generated by a single programmable DAC channel and up-mixed with a locked local oscillator (LO) to drive the klystron preamplifier. The IF frequency must be high enough to meet the system bandwidth requirements and allow filters to be realized which reject the image and the LO leakage. Frequency multipliers can be used to raise the IF frequency without increasing the DAC clock rate at the expense of phase resolution. A single sideband modulator can be used to reduce the image amplitude.

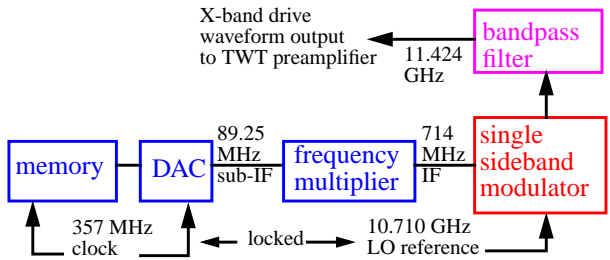

Fig. 3. Digital IF technique for driving pulsed klystrons.

It is important to note that the phase of the output RF is a function of the phase of the (multiplied) DAC produced tone burst and the phase of the LO when the DAC is triggered. By choosing the IF frequency to be an integer multiple of the bunch separation frequency (357 MHz for NLC), the phase of the accelerating RF will repeat for each bunch time slot (equations 1-4). This eliminates the need to load a differently phased DAC waveform or to resynchronize the LO before each machine pulse.

$$\phi_{IF} + \phi_{LO} = \phi_{RF} \qquad [1]$$

$$8\phi_{subIF} + \phi_{LO} = \phi_{RF} \qquad [2]$$

$$8\left(\frac{\pi}{2}T + \phi_{subIF}\right) + (30(2\pi)T + \phi_{LO}) = \phi_{RF} \qquad [3]$$

$$2(2\pi)T + 8\phi_{subIF} + (30(2\pi)T + \phi_{LO}) = \phi_{RF} \qquad [4]$$

Equations 1-4. Derivation showing RF phase repeats every bunch separation (T) interval. T is $(357 \text{ MHz})^{-1}$ for NLC.

To estimate the DAC resolution required to produce 1° drive phase shifts refer to figure 4. If we use the full DAC range to synthesize the IF waveform then the minimum phase shift we can resolve corresponds to a one bit change at a zero crossing. A 7 bit DAC at the IF frequency is required. If multiplication is used to raise the IF frequency more bits are needed. A x8 multiplier requires 3 additional bits. Producing the 89.15 MHz sub-IF with a 12 bit device will allow operating the DAC below full scale.

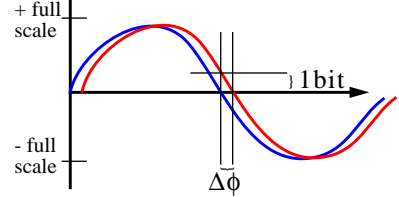

$$(\text{full scale})\sin(\omega t + \Delta\phi) = 1 \text{ count} \qquad [5]$$

At zero crossing:

$$\text{full scale} = \frac{1}{\sin(\Delta\phi)} = \frac{1}{\sin(1°)} = 57 \text{ counts} \qquad [6]$$

$$\text{bits required} = \log_2(2 \cdot 57) \cong 7 \text{ bits at 714 MHz} \qquad [7]$$

Fig. 4. Bits required to achieve 1° phase resolution.

Applying the digital IF signal generation technique to other NLC linac systems operating at different frequencies simply requires a different RF modulator and LO reference. The system bandwidth required to support X-band linac pulse switching will easily support SLED cavity PSK or compressor beam loading compensation requirements. Maintaining the systems will also become less specialized since the DAC/IF hardware and some software will be identical for all klystrons operating at L,C,S, or X-band.

## 3. DIGITAL RF VECTOR DETECTION

To configure the NLC main linac RF systems, measurement techniques must be available to allow proper alignment of the accelerating RF to the beam. Again, a digital IF technique is being planned. The unknown RF signal is mixed down to an IF frequency, amplified, dither added (optional), and sampled with a high speed ADC (figure 5). The choice of the receiver IF frequency is less constrained than for klystron drive generation. The 89.25 MHz IF system shown will have 4 possible phase offsets for measurements triggered on 4 consecutive 357 MHz clocks. If the sample phase offset for each measured IF pulse were known it could be corrected for during post processing. Alternatively, a 357 MHz IF could be used. Undersampling techniques would be used to keep sample rates and memory requirements reasonable while maintaining sufficient channel bandwidth.

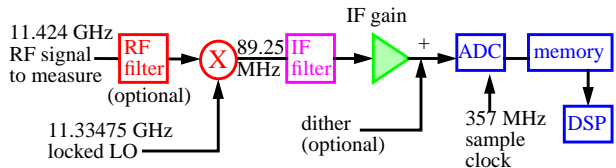

Fig. 5. Digital receiver to accurately measure IF vectors.

Deriving wideband amplitude and phase vectors from a sampled IF waveform can be achieved by applying a Hilbert transform. The Hilbert technique involves taking the Fourier transform of the sampled IF data, nulling all negative frequency bins, scaling positive bins by 2, and finally taking the inverse Fourier transform. This produces a complex time domain vector allowing calculation of amplitude and phase vectors (figure 6). While in the frequency domain, filtering may be applied by nulling any undesired spectral bins, potentially enhancing algorithm efficiency.

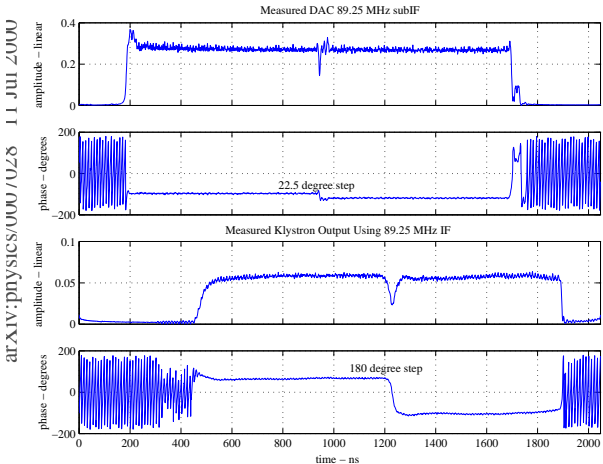

Fig. 6. Driving a klystron and detecting the output using digital IF techniques. The receiver bandwidth was 20 MHz.

To allow automated configuration of the RF systems several measurements must be available (figure 7). The RF output of each klystron output coupler (P1,P2) must be measured to monitor klystron performance. The loaded accelerating RF is measured to allow compensating for beam loading and properly phasing the RF to the beam. Structure transverse alignment is determined by measuring the beam induced dipole modes (X,Y). A beam pickup can provide a phase reference if the receiver LO is chosen not to be locked to the machine reference.

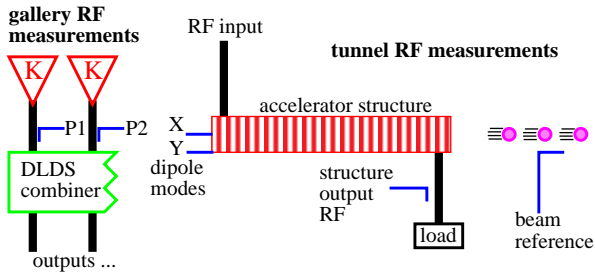

Fig. 7. Measurements needed for automated configuration.

The most difficult measurement task is determining the RF/beam phase during single bunch operation. The structure output RF port allows direct measurement of the accelerating RF, loaded RF, or the beam induced RF (no input RF applied). Dynamic range is an issue since a single pilot bunch of 1e9 particles induces RF 40dB below the accelerating RF. By applying dither to the IF (figure 5) and averaging 10 pulses, the pilot bunch phase can be measured to 1° accuracy (no klystron RF). Alternatively the phase of the accelerating RF can be compared to the loaded RF phase during a single pulse. The vector diagram (figure 8) indicates that the relative angle must be measured to 0.01° accuracy. Simulations show 200 averages of a dithered 89.25 MHz IF using a 7 bit ADC to sample the 285ns burst would be required. Both techniques are truly differential allowing absolute resolution of RF/beam phase and support totally automated linac phasing.

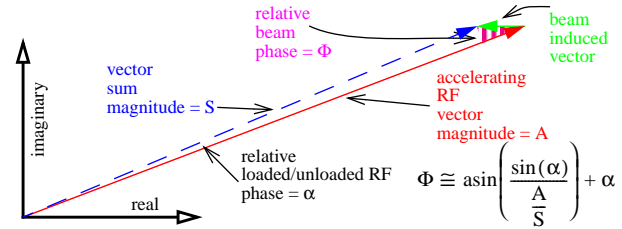

Fig. 8. Vector diagram of RF, beam, and loaded RF. Modern techniques can measure the true RF/beam phase.

Measurement of the 14-16 GHz structure dipole modes requires a variable LO source to mix the desired signal down to the IF frequency. A filter and a RF limiter before the mixer are required to limit peak power when the structure is misaligned. A triple bandpass filter passes dipole modes harmonically related to bunch spacing and corresponding to the center and both ends of the structure. Additional IF gain (30dB) will be applied when the signals become small to produce the 65 dB dynamic range required to measure 1 micron offsets with a single shot pilot bunch (no averaging). Recently a receiver has been tested on a structure and beam reference pickup installed in the SLAC linac (figure 9).

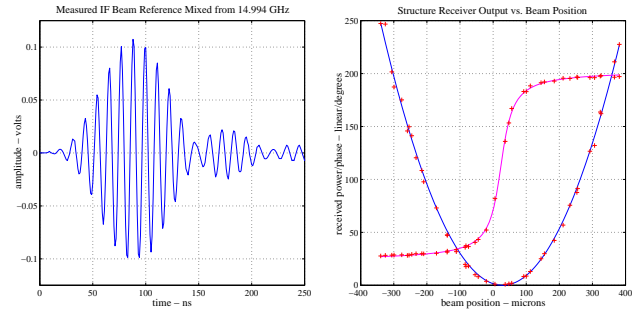

Fig. 9. IF waveform/results from structure alignment test. IF tone burst was digitized using a 1 GHz 8 bit VXI scope.

## 4. CONCLUSION

The digital IF technique proposed to produce arbitrary drive waveforms and accurately detect accelerator RF signals has produced encouraging initial tests results. The development of a high speed DAC/ADC module is planned. A full 8 pack test installation is planned.